\documentclass[a4paper,12pt]{article}

\usepackage{amsmath}
\usepackage{amssymb}
\usepackage{amsthm} 
\usepackage[T1]{fontenc}
\usepackage[bf,small,tableposition=top]{caption}
\usepackage{subfig}
\usepackage{setspace}
 \usepackage{multirow}
\usepackage{graphicx}
\usepackage{epstopdf}
\usepackage{array}
\usepackage{lscape}
\usepackage{booktabs,caption,fixltx2e}
\usepackage[flushleft]{threeparttable}
\usepackage{rotating}
\usepackage{natbib}
\usepackage{mathtools}
\usepackage{stmaryrd}
\usepackage{textcomp}

\theoremstyle{plain}
   \newtheorem{thm}{Theorem}

   \newtheorem{cor}{Corollary}
   
\begin{document}

\begin{center}
 \Large\bf{On simultaneous best linear unbiased prediction of future order statistics and associated properties}
\end{center}
\begin{center}
\bf {\footnotesize  Narayanaswamy Balakrishnan$^1$ and Ritwik Bhattacharya$^2$}
\end{center}
 \begin{center}
\textit{\scriptsize $^1$Department of Mathematics and Statistics, McMaster University, Hamilton, ON L8S 4K1, Canada}\\
 
\textit{\scriptsize $^2$Department of Industrial Engineering, School of Engineering and Sciences, Tecnol\'{o}gico de Monterrey, Quer\'{e}taro 76130,  M\'{e}xico\\}

 \end{center}

\begin{abstract}
In this article, the joint best linear unbiased predictors (BLUPs) of two future unobserved order statistics, based on a set of observed order statistics, are developed explicitly. It is shown that these predictors are trace-efficient as well as determinant-efficient BLUPs. More generally, the BLUPs are shown to possess complete mean squared predictive error matrix dominance in the class of all linear unbiased predictors of two future unobserved order statistics. Finally, these results are extended to the case of simultaneous BLUPs of any $l$ future order statistics.
\end{abstract}

 {\textbf{Keywords}}:  Best linear unbiased predictor, Best linear unbiased estimator, Order statistics, Trace-efficient predictor, Determinant-efficient predictor, Mean squared predictive error, Complete mean squared predictive error matrix dominance.

\section{Introduction} \label{sec1}
\paragraph{}
The issue of prediction of future unobserved failure times has been of great interest in reliability life-testing experiments. The problem can be mathematically formulated as follows. Consider a continuous (lifetime) distribution with probability density function
\begin{equation}\label{pdf}
	\frac{1}{\sigma} f\left(\frac{x-\mu}{\sigma} \right),
\end{equation}
where $\mu$ and $\sigma$ are the location and scale parameters, respectively. Suppose the first $r$ order statistics (that is, a Type-II right censored sample) $$X_{1:n}<X_{2:n}<\cdots<X_{r:n},$$ out of a sample of size $n$ from (\ref{pdf}), are observed. We then need to predict the unobserved future order statistics $X_{r+1:n}, X_{r+2:n},\cdots, X_{n:n}$ based on the $r$ observed order statistics. \\

To begin with, let us focus on the joint prediction of $X_{s:n}$ and $X_{t:n},$ where $r<s<t\leq n$. Let us use the following notation: $\boldsymbol{X} = (X_{1:n}, \cdots, X_{r:n})^{\prime}_{r\times1},$ $\alpha_i$ for the expected value of the standardized order statistic $Z_{i:n}=(X_{i:n}-\mu)/\sigma, ~i=1,\cdots,n,$ with $\boldsymbol{\alpha}=(\alpha_1, \cdots, \alpha_r)^{\prime}_{r\times1},$ and $\sigma^2 \boldsymbol{\Sigma}$ for the variance-covariance matrix of $\boldsymbol{X}$, where $ \boldsymbol{\Sigma}$ is the $r\times r$ covariance matrix of $Z_{i:n},  i = 1,\cdots, r$, assumed to be positive definite. Using these notation, the marginal best linear unbiased predictor $\hat{X}_{s:n}$ of $X_{s:n}$ has been derived by \cite{Kaminsky_1975}, using the earlier results of \cite{Goldberger_1962},  as
\begin{equation}\label{BLUPs}
	\hat{X}_{s:n} = \hat{\mu} + \hat{\sigma}\alpha_s + \boldsymbol{\omega}^{\prime}_s\boldsymbol{\Sigma}^{-1}(\boldsymbol{X}-\hat{\mu}\boldsymbol{1} - \hat{\sigma}\boldsymbol{\alpha}),
\end{equation}where $\boldsymbol{1} = (1,\cdots, 1)^{\prime}_{r\times 1}$ and $\boldsymbol{\omega}_s = (\omega_1,\cdots, \omega_r)^{\prime}_{r\times 1}$, with $\omega_i = \mbox{Cov}(Z_{i:n}, Z_{s:n})$.  Similarly, the marginal best linear unbiased predictor $\hat{X}_{t:n}$ of $X_{t:n}$ is then exactly as in (\ref{BLUPs}) with $\alpha_s$ and $\boldsymbol{\omega}^{\prime}_s$ being replaced by $\alpha_t$ and $\boldsymbol{\omega}^{\prime}_t$, respectively. In (\ref{BLUPs}), $\hat{\mu}$ and $\hat{\sigma}$ are the best linear unbiased estimators (BLUEs) of $\mu$ and $\sigma$, respectively, based on the observed $\boldsymbol{X} $, given by
{\small{\begin{eqnarray}\nonumber
			\hat{\mu} &=& \frac{1}{\Delta}\{ (\boldsymbol{\alpha}^{\prime} \boldsymbol{\Sigma}^{-1} \boldsymbol{\alpha})(\boldsymbol{1}^{\prime} \boldsymbol{\Sigma}^{-1}) -  (\boldsymbol{\alpha}^{\prime} \boldsymbol{\Sigma}^{-1} \boldsymbol{1})(\boldsymbol{\alpha}^{\prime} \boldsymbol{\Sigma}^{-1})\}\boldsymbol{X},\\\nonumber
			\hat{\sigma} &=& \frac{1}{\Delta}\{ (\boldsymbol{1}^{\prime} \boldsymbol{\Sigma}^{-1} \boldsymbol{1})(\boldsymbol{\alpha}^{\prime} \boldsymbol{\Sigma}^{-1}) -  (\boldsymbol{1}^{\prime} \boldsymbol{\Sigma}^{-1} \boldsymbol{\alpha})(\boldsymbol{1}^{\prime} \boldsymbol{\Sigma}^{-1})\}\boldsymbol{X},
		\end{eqnarray}
	with
	\begin{eqnarray}\nonumber
			\mbox{Var}(	\hat{\mu}) &=& \frac{\boldsymbol{\alpha}^{\prime} \boldsymbol{\Sigma}^{-1} \boldsymbol{\alpha}}{\Delta} \sigma^2,~ \mbox{Var}(	\hat{\sigma}) ~=~ \frac{\boldsymbol{1}^{\prime} \boldsymbol{\Sigma}^{-1} \boldsymbol{1}}{\Delta}\sigma^2,~ \mbox{Cov}(	\hat{\mu},	\hat{\sigma}) ~=~ -\frac{\boldsymbol{1}^{\prime} \boldsymbol{\Sigma}^{-1} \boldsymbol{\alpha}}{\Delta}\sigma^2
	\end{eqnarray}	
and
	\begin{eqnarray}		\label{bigdelta}
			\Delta &=& (\boldsymbol{1}^{\prime} \boldsymbol{\Sigma}^{-1} \boldsymbol{1}) (\boldsymbol{\alpha}^{\prime} \boldsymbol{\Sigma}^{-1} \boldsymbol{\alpha}) - (\boldsymbol{1}^{\prime} \boldsymbol{\Sigma}^{-1} \boldsymbol{\alpha})^2,
\end{eqnarray}}}being the determinant of the matrix
	\begin{equation*}
		\begin{bmatrix} \boldsymbol{1}^{\prime}\boldsymbol{\Sigma}^{-1}\boldsymbol{1} &  \boldsymbol{1}^{\prime}\boldsymbol{\Sigma}^{-1}\boldsymbol{\alpha}\\ \boldsymbol{1}^{\prime}\boldsymbol{\Sigma}^{-1}\boldsymbol{\alpha} & \boldsymbol{\alpha}^{\prime}\boldsymbol{\Sigma}^{-1}\boldsymbol{\alpha}\end{bmatrix}, 
	\end{equation*}which is related to the variance-covariance matrix of the BLUEs $(\hat{\mu}, \hat{\sigma})$; see \cite{Cohen_book} and \cite{Nagaraja_book} for pertinent details. Note that the predictors $\hat{X}_{s:n}$ and $\hat{X}_{t:n}$ are indeed trace-efficient predictors by their very construction.  It is important to mention that the above mentioned BLUEs ($\hat{\mu}$ and $\hat{\sigma}$) and BLUP ($\hat{X}_{s:n}$) possess several interesting properties. Interested readers may refer to \cite{Rao_1997}, \cite{Doganaksoy_1997} and \cite{Rao_book1, Bala_2003} for details. Recently, \cite{Bala_2021} derived explicit expressions for the joint BLUPs of two future order statistics by minimizing the determinant of the variance–covariance matrix of the predictors, and also showed the non-existence of joint BLUPs when three or more order statistics are considered. \\

In this article, we first derive explicit expressions of joint BLUPs obtained by minimizing the determinant of the mean squared predictive error matrix. Thence, we establish the property that the BLUPs are also determinant-efficient predictors. More generally, we then establish the complete mean squared predictive error matrix dominance property of these joint BLUPs, similar in principle to the complete covariance matrix dominance property of BLUEs established by \cite{Rao_1997, Bala_2003}. Finally, we extend these results to the general case of simultaneous best linear unbiased prediction of any $l$ future order statistics.

\section{Determinant-efficiency of BLUPs}
\paragraph{}
In this section, we first derive explicit expressions for the joint predictors of two future order statistics and the corresponding mean squared predictive error matrix, as presented in the following theorem.

\begin{thm}
The joint best linear unbiased predictors $\tilde{X}_{s:n}$ and $\tilde{X}_{t:n}$, determined by minimizing the determinant of the mean squared predictive error matrix, are of the form $\tilde{X}_{s:n} = \boldsymbol{a}^{\prime}\boldsymbol{X}$ and $\tilde{X}_{t:n} = \boldsymbol{b}^{\prime}\boldsymbol{X}$ in which the coefficients $\boldsymbol{a}=(a_1, \cdots, a_r)^{\prime}_{r\times 1}$ and $\boldsymbol{b}=(b_1, \cdots, b_r)^{\prime}_{r\times 1}$ are given by	
\begin{equation*}
	\boldsymbol{a} = \boldsymbol{\Sigma}^{-1}\boldsymbol{\omega}_s + \frac{1}{\Delta} (V_2A_s - V_3B_s)\boldsymbol{\Sigma}^{-1}\boldsymbol{1} +\frac{1}{\Delta} (V_1B_s - V_3A_s) \boldsymbol{\Sigma}^{-1}\boldsymbol{\alpha}
\end{equation*}	
	and
	\begin{equation*}
		\boldsymbol{b} = \boldsymbol{\Sigma}^{-1}\boldsymbol{\omega}_t + \frac{1}{\Delta} (V_2A_t - V_3B_t)\boldsymbol{\Sigma}^{-1}\boldsymbol{1} +\frac{1}{\Delta} (V_1B_t - V_3A_t) \boldsymbol{\Sigma}^{-1}\boldsymbol{\alpha},
	\end{equation*}where $V_1 = \boldsymbol{1}^{\prime}\boldsymbol{\Sigma}^{-1}\boldsymbol{1}$, $V_2 = \boldsymbol{\alpha}^{\prime}\boldsymbol{\Sigma}^{-1}\boldsymbol{\alpha}$, $V_3=\boldsymbol{1}^{\prime}\boldsymbol{\Sigma}^{-1}\boldsymbol{\alpha}$, $A_s = 1-\boldsymbol{1}^{\prime}\boldsymbol{\Sigma}^{-1}\boldsymbol{\omega}_s$, $A_t = 1-\boldsymbol{1}^{\prime}\boldsymbol{\Sigma}^{-1}\boldsymbol{\omega}_t$, $B_s = \alpha_s-\boldsymbol{\alpha}^{\prime}\boldsymbol{\Sigma}^{-1}\boldsymbol{\omega}_s$, $B_t = \alpha_t-\boldsymbol{\alpha}^{\prime}\boldsymbol{\Sigma}^{-1}\boldsymbol{\omega}_t$ and $\Delta$ is as defined earlier in (\ref{bigdelta}).
\end{thm}	  

\noindent\textbf{Proof:} For deriving the joint BLUPs, obtained by minimizing the determinant of the mean squared predictive error (MSPE) matrix of the predictors, we consider the MSPE matrix given by
\begin{equation*}
 \sigma^2 \begin{bmatrix} W_{1} &  W_{3}\\ W_{3} & W_{2}\end{bmatrix},
\end{equation*}where 
\begin{eqnarray}\nonumber
	W_1 &=& \boldsymbol{a}^{\prime}\boldsymbol{\Sigma}\boldsymbol{a} -2\boldsymbol{a}^{\prime}\boldsymbol{\omega}_s - \omega_{ss},\\\label{WVar}
	W_2 &=& \boldsymbol{b}^{\prime}\boldsymbol{\Sigma}\boldsymbol{b} -2\boldsymbol{b}^{\prime}\boldsymbol{\omega}_t - \omega_{tt},\\\nonumber
	W_3 &=& \boldsymbol{a}^{\prime}\boldsymbol{\Sigma}\boldsymbol{b} -\boldsymbol{a}^{\prime}\boldsymbol{\omega}_t - \boldsymbol{b}^{\prime}\boldsymbol{\omega}_s - \omega_{st},
\end{eqnarray}with $\omega_{ss}$, $\omega_{tt}$ and $\omega_{st}$ being $\mbox{Var}(Z_{s:n})$, $\mbox{Var}(Z_{t:n})$ and $\mbox{Cov}(Z_{s:n}, Z_{t:n})$, respectively. In order for the obtained predictors to be unbiased, we need to impose the following four constraints: $\boldsymbol{a}^{\prime}\boldsymbol{1}=1$, $\boldsymbol{a}^{\prime}\boldsymbol{\alpha}=\alpha_s$, $\boldsymbol{b}^{\prime}\boldsymbol{1}=1$ and $\boldsymbol{b}^{\prime}\boldsymbol{\alpha}=\alpha_t$. Thus, for the required constrained optimization, we shall employ the Lagrangian method to determine $\boldsymbol{a}$ and $\boldsymbol{b}$ optimally by minimizing the objective function (ignoring the multiplicative factor $\sigma^2$)
\begin{eqnarray}\nonumber
	Q(\boldsymbol{a}, \boldsymbol{b}) &=& W_1W_2-W_3^2 - 2\lambda_1(\boldsymbol{a}^{\prime}\boldsymbol{1}-1) - 2\lambda^{*}_1(\boldsymbol{a}^{\prime}\boldsymbol{\alpha}-\alpha_s)\\\label{Lobg}
	&&-2\lambda_2(\boldsymbol{b}^{\prime}\boldsymbol{1}-1) - 2\lambda^{*}_2(\boldsymbol{b}^{\prime}\boldsymbol{\alpha}-\alpha_t).
\end{eqnarray}
Upon differentiating  (\ref{Lobg}) with respect to $ \boldsymbol{a}$ and $ \boldsymbol{b}$ and equating them to vector $ \boldsymbol{0}$ of dimension $r\times 1$, we obtain 
\begin{eqnarray}\label{E1}
	\boldsymbol{a} &=& \boldsymbol{\Sigma}^{-1}\boldsymbol{\omega}_s + \frac{W_3}{W_2}(\boldsymbol{b} -  \boldsymbol{\Sigma}^{-1}\boldsymbol{\omega}_t)+\frac{\lambda_1}{W_2}\boldsymbol{\Sigma}^{-1}\boldsymbol{1}+\frac{\lambda_1^*}{W_2}\boldsymbol{\Sigma}^{-1}\boldsymbol{\alpha}, \\\label{E2}
	\boldsymbol{b} &=& \boldsymbol{\Sigma}^{-1}\boldsymbol{\omega}_t + \frac{W_3}{W_1}(\boldsymbol{a} -  \boldsymbol{\Sigma}^{-1}\boldsymbol{\omega}_s)+\frac{\lambda_2}{W_1}\boldsymbol{\Sigma}^{-1}\boldsymbol{1}+\frac{\lambda_2^*}{W_1}\boldsymbol{\Sigma}^{-1}\boldsymbol{\alpha},
\end{eqnarray}respectively. Now, solving (\ref{E1}) and (\ref{E2}) for $\boldsymbol{a}$ and $\boldsymbol{b}$, we obtain
\begin{eqnarray}\nonumber
	\boldsymbol{a} &=& \boldsymbol{\Sigma}^{-1}\boldsymbol{\omega}_s + \frac{W_1}{\Delta_1} \lambda_1\boldsymbol{\Sigma}^{-1}\boldsymbol{1} + \frac{W_1}{\Delta_1} \lambda_1^*\boldsymbol{\Sigma}^{-1}\boldsymbol{\alpha}\\\label{E7}
	&& +\frac{W_3}{\Delta_1} \lambda_2\boldsymbol{\Sigma}^{-1}\boldsymbol{1} + \frac{W_3}{\Delta_1} \lambda_2^* \boldsymbol{\Sigma}^{-1}\boldsymbol{\alpha},\\\nonumber
	\boldsymbol{b} &=& \boldsymbol{\Sigma}^{-1}\boldsymbol{\omega}_t + \frac{W_3}{\Delta_1} \lambda_1\boldsymbol{\Sigma}^{-1}\boldsymbol{1} + \frac{W_2}{\Delta_1} \lambda_1^*\boldsymbol{\Sigma}^{-1}\boldsymbol{\alpha}\\\label{E6}
&& +\frac{W_3}{\Delta_1} \lambda_2\boldsymbol{\Sigma}^{-1}\boldsymbol{1} + \frac{W_2}{\Delta_1} \lambda_2^* \boldsymbol{\Sigma}^{-1}\boldsymbol{\alpha},
\end{eqnarray}where $\Delta_1 = W_1W_2-W_3^2$ is assumed to be positive.\\

Next, applying the unbiasedness conditions  $\boldsymbol{a}^{\prime}\boldsymbol{1}=1$, $\boldsymbol{a}^{\prime}\boldsymbol{\alpha}=\alpha_s$, $\boldsymbol{b}^{\prime}\boldsymbol{1}=1$ and $\boldsymbol{b}^{\prime}\boldsymbol{\alpha}=\alpha_t$ in (\ref{E7}) and (\ref{E6}), and simplifying the resulting equations using the notation introduced earlier, we get
\begin{eqnarray*}
	\lambda_1 &=& \frac{1}{\Delta}\left[  W_2(V_2A_s - V_3B_s) - W_3(V_2A_t - V_3B_t)    \right],\\
	\lambda_2 &=&  \frac{1}{\Delta}\left[  W_1(V_2A_t - V_3B_t) - W_3(V_2A_s - V_3B_s)    \right],\\
	\lambda_1^* &=& \frac{1}{\Delta} \left[  W_2(V_1B_s - V_3A_s) - W_3(V_1B_t - V_3A_t)   \right],\\
	\lambda_2^* &=& \frac{1}{\Delta} \left[  W_1(V_1B_t - V_3A_t) - W_3(V_1B_s - V_3A_s)   \right].
\end{eqnarray*}Finally, upon substituting the above determined optimal values of $\lambda_1$, $\lambda_2$, $\lambda_1^*$ and $\lambda_2^*$ in (\ref{E7}) and (\ref{E6}) and simplifying, we obtain the desired coefficient vectors of the joint BLUPs to be
\begin{equation*}
	\boldsymbol{a} = \boldsymbol{\Sigma}^{-1}\boldsymbol{\omega}_s + \frac{1}{\Delta} (V_2A_s - V_3B_s)\boldsymbol{\Sigma}^{-1}\boldsymbol{1} +\frac{1}{\Delta} (V_1B_s - V_3A_s) \boldsymbol{\Sigma}^{-1}\boldsymbol{\alpha}
\end{equation*}	
and
\begin{equation*}
	\boldsymbol{b} = \boldsymbol{\Sigma}^{-1}\boldsymbol{\omega}_t + \frac{1}{\Delta} (V_2A_t - V_3B_t)\boldsymbol{\Sigma}^{-1}\boldsymbol{1} +\frac{1}{\Delta} (V_1B_t - V_3A_t) \boldsymbol{\Sigma}^{-1}\boldsymbol{\alpha}.
\end{equation*}Hence, the theorem.

\begin{cor}
	Upon substituting the above expressions of $\boldsymbol{a}$ and $\boldsymbol{b}$ in the expressions of $W_1$, $W_2$ and $W_3$ in (\ref{WVar}), and going through some lengthy algebraic  calculations, we derive the MSPE matrix of the joint BLUPs to be
	\begin{equation*}
		\sigma^2 \begin{bmatrix} W_{1} &  W_{3}\\ W_{3} & W_{2}\end{bmatrix},
	\end{equation*}where 
	\begin{eqnarray*}
		W_1 &=& \omega_{ss} - \boldsymbol{\omega}^{'}_s\boldsymbol{\Sigma}^{-1}\boldsymbol{\omega}_s + \begin{bmatrix} A_s &  B_s\end{bmatrix}\begin{bmatrix} V_{1} &  V_{3}\\ V_{3} & V_{2}\end{bmatrix}^{-1}\begin{bmatrix} A_s\\ B_s \end{bmatrix},\\
		W_2 &=& \omega_{tt} - \boldsymbol{\omega}^{'}_t\boldsymbol{\Sigma}^{-1}\boldsymbol{\omega}_t + \begin{bmatrix} A_t &  B_t\end{bmatrix}\begin{bmatrix} V_{1} &  V_{3}\\ V_{3} & V_{2}\end{bmatrix}^{-1}\begin{bmatrix} A_t\\ B_t \end{bmatrix},\\
		W_3 &=& \omega_{st} - \boldsymbol{\omega}^{'}_s\boldsymbol{\Sigma}^{-1}\boldsymbol{\omega}_t + \begin{bmatrix} A_s &  B_s\end{bmatrix}\begin{bmatrix} V_{1} &  V_{3}\\ V_{3} & V_{2}\end{bmatrix}^{-1}\begin{bmatrix} A_t\\ B_t \end{bmatrix}.
	\end{eqnarray*}
\end{cor}	

\subsection{Equivalence of joint BLUPs and marginal BLUPs}
\paragraph{}
We now establish the equivalence of joint BLUPs and marginal BLUPs. 
\begin{thm}
	The joint BLUPs $\tilde{X}_{s:n}$ and $\tilde{X}_{t:n}$ derived in Theorem 1 and the marginal BLUPs $\hat{X}_{s:n}$ and $\hat{X}_{t:n}$ presented in (\ref{BLUPs}) are equivalent. 
\end{thm}

\noindent\textbf{Proof:} To prove the required result, we take the expressions of joint BLUPs and then simplify them to show that they are indeed the marginal BLUPs. For this purpose, let us write from Theorem 1, 
\begin{eqnarray*}
\tilde{X}_{s:n}&=& 	\boldsymbol{a}^{\prime}\boldsymbol{X}\\
&=& \boldsymbol{\omega}^{\prime}_s\boldsymbol{\Sigma}^{-1}\boldsymbol{X} + \frac{1}{\Delta} (V_2A_s - V_3B_s)\boldsymbol{1}^{\prime}\boldsymbol{\Sigma}^{-1}\boldsymbol{X} +\frac{1}{\Delta} (V_1B_s - V_3A_s) \boldsymbol{\alpha}^{\prime}\boldsymbol{\Sigma}^{-1}\boldsymbol{X}\\
&=&\boldsymbol{\omega}^{\prime}_s\boldsymbol{\Sigma}^{-1}\boldsymbol{X} \\
&&+ \frac{1}{\Delta} [ (\boldsymbol{\alpha}^{\prime} \boldsymbol{\Sigma}^{-1} \boldsymbol{\alpha})\{1-(\boldsymbol{1}^{\prime} \boldsymbol{\Sigma}^{-1} \boldsymbol{\omega}_s)\}-(\boldsymbol{1}^{\prime} \boldsymbol{\Sigma}^{-1} \boldsymbol{\alpha})\{\alpha_s - (\boldsymbol{\alpha}^{\prime} \boldsymbol{\Sigma}^{-1} \boldsymbol{\omega}_s) \}](\boldsymbol{1}^{\prime}\boldsymbol{\Sigma}^{-1}\boldsymbol{X})\\
&&+\frac{1}{\Delta} [ (\boldsymbol{1}^{\prime} \boldsymbol{\Sigma}^{-1} \boldsymbol{1})\{\alpha_s  - (\boldsymbol{\alpha}^{\prime} \boldsymbol{\Sigma}^{-1} \boldsymbol{\omega}_s)  \}- (\boldsymbol{1}^{\prime} \boldsymbol{\Sigma}^{-1} \boldsymbol{\alpha})\{1- (\boldsymbol{1}^{\prime} \boldsymbol{\Sigma}^{-1} \boldsymbol{\omega}_s)   \}]  (\boldsymbol{\alpha}^{\prime} \boldsymbol{\Sigma}^{-1} X)
\end{eqnarray*}	
\begin{eqnarray*}
	&=&\boldsymbol{\omega}^{\prime}_s\boldsymbol{\Sigma}^{-1}\boldsymbol{X} \\
	&&+\frac{1}{\Delta} [(\boldsymbol{\alpha}^{\prime} \boldsymbol{\Sigma}^{-1} \boldsymbol{\alpha})(\boldsymbol{1}^{\prime} \boldsymbol{\Sigma}^{-1}) -  (\boldsymbol{\alpha}^{\prime} \boldsymbol{\Sigma}^{-1} \boldsymbol{1})(\boldsymbol{\alpha}^{\prime} \boldsymbol{\Sigma}^{-1}) ]\boldsymbol{X}\\
	&&+ \alpha_s \frac{1}{\Delta} [ (\boldsymbol{1}^{\prime} \boldsymbol{\Sigma}^{-1} \boldsymbol{1})(\boldsymbol{\alpha}^{\prime} \boldsymbol{\Sigma}^{-1}) -  (\boldsymbol{1}^{\prime} \boldsymbol{\Sigma}^{-1} \boldsymbol{\alpha})(\boldsymbol{1}^{\prime} \boldsymbol{\Sigma}^{-1})]\boldsymbol{X}\\
	&&+ \boldsymbol{\omega}^{\prime}_s\boldsymbol{\Sigma}^{-1}  \frac{1}{\Delta} [ -\boldsymbol{1} \{ (\boldsymbol{\alpha}^{\prime} \boldsymbol{\Sigma}^{-1} \boldsymbol{\alpha})(\boldsymbol{1}^{\prime} \boldsymbol{\Sigma}^{-1}) -  (\boldsymbol{\alpha}^{\prime} \boldsymbol{\Sigma}^{-1} \boldsymbol{1})(\boldsymbol{\alpha}^{\prime} \boldsymbol{\Sigma}^{-1})\}\boldsymbol{X}\\
	&& ~~~~~~~~~~~~~~~~~~~~~~~-\boldsymbol{\alpha}\{ (\boldsymbol{1}^{\prime} \boldsymbol{\Sigma}^{-1} \boldsymbol{1})(\boldsymbol{\alpha}^{\prime} \boldsymbol{\Sigma}^{-1}) -  (\boldsymbol{1}^{\prime} \boldsymbol{\Sigma}^{-1} \boldsymbol{\alpha})(\boldsymbol{1}^{\prime} \boldsymbol{\Sigma}^{-1})\}\boldsymbol{X}]\\
	&=& \boldsymbol{\omega}^{\prime}_s\boldsymbol{\Sigma}^{-1}\boldsymbol{X} +\hat{\mu} + \alpha_s\hat{\sigma } + \boldsymbol{\omega}^{\prime}_s\boldsymbol{\Sigma}^{-1}(-\hat{\mu}\boldsymbol{1} - \hat{\sigma}\boldsymbol{\alpha})\\
	&=& \hat{\mu} + \alpha_s\hat{\sigma } + \boldsymbol{\omega}^{\prime}_s\boldsymbol{\Sigma}^{-1}(\boldsymbol{X}-\hat{\mu}\boldsymbol{1} - \hat{\sigma}\boldsymbol{\alpha})\\
	&=& \hat{X}_{s:n}.
\end{eqnarray*}In a similar manner, we can show that $\tilde{X}_{t:n}=\hat{X}_{t:n}$. Hence, the theorem. 

\section{Complete MSPE matrix dominance of BLUPs}
\paragraph{}
Consider the problem of predicting the random quantity $Y = lX_{s:n} + kX_{t:n},$ which is a linear combination of two future order statistics $X_{s:n}$ and $X_{t:n}$, with $l$ and $k$ being two fixed constants. Now, let us consider an arbitrary linear predictor for $Y$ as $$\hat{Y}=\boldsymbol{c}^{\prime}\boldsymbol{X},$$where the coefficient vector $\boldsymbol{c}=(c_1, \cdots, c_r)^{\prime}_{r\times 1}$ needs to be suitably determined. Note that, in order for the predictor $\hat{Y}$ to be unbiased for $Y$, the following two unbiasedness conditions 
\begin{equation}\label{LM}
	\boldsymbol{c}^{\prime}\boldsymbol{1} = l+k \mbox{ and } \boldsymbol{c}^{\prime}\boldsymbol{\alpha} = l\alpha_s +k\alpha_t
\end{equation}need to be satisfied. As the mean squared predictive error of $\hat{Y}$ is given by 
\begin{equation*}
	W = \sigma^2 (\boldsymbol{c}^{\prime}\boldsymbol{\Sigma}\boldsymbol{c} - 2l\boldsymbol{c}^{\prime}\boldsymbol{\omega}_s - 2k\boldsymbol{c}^{\prime}\boldsymbol{\omega}_t + l^2\omega_{ss} + k^2 \omega_{tt} + lk \omega_{st}),
\end{equation*}we need to minimize $W$ subject to the unbiasedness conditions for determining the BLUP for $Y$. Again applying the Lagrangian optimization method, the optimal solution for $\boldsymbol{c}$ can be found to be 
\begin{equation}\label{C1}
	\boldsymbol{c} = l \boldsymbol{\Sigma}^{-1}\boldsymbol{\omega}_s  + k \boldsymbol{\Sigma}^{-1}\boldsymbol{\omega}_t + \lambda_1 \boldsymbol{\Sigma}^{-1}\boldsymbol{1}+ \lambda_2 \boldsymbol{\Sigma}^{-1}\boldsymbol{\alpha}.
\end{equation}
 Using the above expression of $\boldsymbol{c}$  in the unbiasedness conditions in (\ref{LM}) and simplifying, we obtain 
 \begin{eqnarray*}
 	\lambda_1 &=& \frac{1}{\Delta} \bigg[V_2(l+k - l \boldsymbol{1}^{\prime}\boldsymbol{\Sigma}^{-1}\boldsymbol{\omega}_s -k \boldsymbol{1}^{\prime}\boldsymbol{\Sigma}^{-1}\boldsymbol{\omega}_t) \\
 	&& -   V_3(l\alpha_s + k \alpha_t - l \boldsymbol{\alpha}^{\prime}\boldsymbol{\Sigma}^{-1}\boldsymbol{\omega}_s  - k \boldsymbol{\alpha}^{\prime}\boldsymbol{\Sigma}^{-1}\boldsymbol{\omega}_t )\bigg],\\
 	\lambda_2 &=& \frac{1}{\Delta}\bigg[V_3(l+k - l \boldsymbol{1}^{\prime}\boldsymbol{\Sigma}^{-1}\boldsymbol{\omega}_s -k \boldsymbol{1}^{\prime}\boldsymbol{\Sigma}^{-1}\boldsymbol{\omega}_t) \\
 	&&+ V_3(l\alpha_s + k \alpha_t - l \boldsymbol{\alpha}^{\prime}\boldsymbol{\Sigma}^{-1}\boldsymbol{\omega}_s  - k \boldsymbol{\alpha}^{\prime}\boldsymbol{\Sigma}^{-1}\boldsymbol{\omega}_t )\bigg].
 \end{eqnarray*}Finally, substituting these expressions of $\lambda_1$ and $\lambda_2$ in (\ref{C1}) and simplifying, we readily obtain $$\boldsymbol{c} = l\boldsymbol{a} + k \boldsymbol{b},$$and consequently, $$\hat{Y} = l \tilde{X}_{s:n} + k\tilde{X}_{t:n},$$where $\tilde{X}_{s:n}$ and $\tilde{X}_{s¡t:n}$ are the joint BLUPs of $X_{s:n}$ and $X_{t:n}$, respectively, based on $\boldsymbol{X}$. As a result, we have $$\mbox{Var}( l \tilde{X}_{s:n} + k\tilde{X}_{t:n})\leq \mbox{Var}( l X^{*}_{s:n} + kX^{*}_{t:n}),$$for any other joint linear unbiased predictors $X^{*}_{s:n}$ and $X^{*}_{t:n}$ of $X_{s:n}$ and $X_{t:n}$. This means 
\begin{eqnarray*}
	&&\begin{bmatrix} l &  k\end{bmatrix}\begin{bmatrix} \mbox{MVar}(\tilde{X}_{s:n}) &  \mbox{MCov}(\tilde{X}_{s:n}, \tilde{X}_{t:n})\\ \mbox{MCov}(\tilde{X}_{s:n}, \tilde{X}_{t:n}) & \mbox{MVar}(\tilde{X}_{t:n})\end{bmatrix}\begin{bmatrix} l \\ k\end{bmatrix}\\
	&&\leq \begin{bmatrix} l &  k\end{bmatrix}\begin{bmatrix} \mbox{MVar}(X^{*}_{s:n}) &  \mbox{MCov}(X^{*}_{s:n}, X^{*}_{t:n})\\ \mbox{MCov}(X^{*}_{s:n}, X^{*}_{t:n}) & \mbox{MVar}(X^{*}_{t:n})\end{bmatrix}\begin{bmatrix} l \\ k\end{bmatrix},
\end{eqnarray*}where MVar and MCov are abbreviations for mean squared predictive error variance and mean squared predictive error covariance, respectively. This establishes the property that the BLUPs of $X_{s:n}$ and $X_{t:n}$ possess complete MSPE matrix dominance in the class of all linear unbiased predictors of $X_{s:n}$ and $X_{t:n}$, which is indeed a more general result than that of trace-efficiency and determinant-efficiency.

\section{Extension to simultaneous BLUPs of $l$ future order statistics}
\paragraph{}	
We now consider the BLUPs of $l$ future order statistics $(X_{s_1:n}, X_{s_2:n},\cdots, X_{s_l:n}),$ for $r<s_1<s_2<\cdots<s_l\leq n,$ simultaneously. We then have the following important property.
\begin{thm}
	The simultaneous best linear unbiased predictors of $l$ future order statistics are identical to their corresponding marginal predictors. 
\end{thm}
\noindent\textbf{Proof:} Let us assume that the BLUPs of $l$ future order statistics are of the form 
\begin{equation}\label{BLUPai}
	\tilde{X}_{s_i:n} = \boldsymbol{a}^{\prime}_i \boldsymbol{X}, ~i = 1, 2,\cdots,l,
\end{equation}where $\boldsymbol{a}^{,}_i$s are coefficient vectors of size $r\times1$ that need to be suitably determined.  The corresponding mean squared predictive error matrix is then $$\sigma^2\boldsymbol{W} =\sigma^2 \big(\big(W_{ij}\big)\big)_{i, j=1}^l,$$where
\begin{equation*}
	W_{ii} = \boldsymbol{a}^{\prime}_i  \boldsymbol{\Sigma} \boldsymbol{a}_i - 2\boldsymbol{a}^{\prime}_i\boldsymbol{\omega}_{s_i} + \omega_{s_is_i},~ i=1, 2,\cdots, l,
\end{equation*}and
\begin{equation*}
	W_{ij} = \boldsymbol{a}^{\prime}_i  \boldsymbol{\Sigma} \boldsymbol{a}_j - \boldsymbol{a}^{\prime}_i\boldsymbol{\omega}_{s_j} - \boldsymbol{a}^{\prime}_j\boldsymbol{\omega}_{s_i} +  \omega_{s_is_i}, ~1\leq i < j\leq l.
\end{equation*}Note that $\boldsymbol{W}$ is symmetric, i.e., $W_{ij} = W_{ji}$. Also, observe that each coefficient vector $\boldsymbol{a}_i$ appears in only one row and column. For instance, $\boldsymbol{a}_i$ appears only in the $i$th row and the $i$th column. Let us use $\sigma^2|\boldsymbol{W}|$ and $C^{ij}$ to denote the determinant and the co-factor of $W_{ij}$, respectively.  Let us further denote 
\begin{eqnarray*}
	\frac{\partial}{\partial \boldsymbol{a}_i } W_{ii} &=& 2( \boldsymbol{\Sigma} \boldsymbol{a}_i -\boldsymbol{\omega}_{s_i}) ~ ~ = ~ ~ W_{ii}^{(i)},\\
	\frac{\partial}{\partial \boldsymbol{a}_i } W_{ij} &=& \boldsymbol{\Sigma} \boldsymbol{a}_j -\boldsymbol{\omega}_{s_j} ~ ~ = ~ ~ \frac{1}{2}W_{jj}^{(j)},\\
	\frac{\partial}{\partial \boldsymbol{a}_j } W_{ij} &=& \boldsymbol{\Sigma} \boldsymbol{a}_i -\boldsymbol{\omega}_{s_i} ~ ~ = ~ ~ \frac{1}{2}W_{ii}^{(i)}.
\end{eqnarray*}Then, to obtain the BLUP in (\ref{BLUPai}), we need to impose the unbiasedness conditions
\begin{eqnarray*}\label{UC1}
	\boldsymbol{a}^{\prime}_i	\boldsymbol{1} &=& 1, ~i = 1, 2,\cdots,l,\\\label{UC2}
	\boldsymbol{a}^{\prime}_i	\boldsymbol{\alpha} &=& \alpha_{s_i}, ~i = 1, 2,\cdots,l.
\end{eqnarray*}We again apply the Lagrangian optimization method to minimize the objective function (ignoring $\sigma^2$ term)
\begin{equation}\label{Lagran}
	L = |\boldsymbol{W}| - 2\sum_{i=1}^l \lambda_i(\boldsymbol{a}^{\prime}_i	\boldsymbol{1}-1)-2\sum_{i=1}^l \lambda^*_i(\boldsymbol{a}^{\prime}_i	\boldsymbol{\alpha} - \alpha_{s_i}),
\end{equation}where $2\lambda_i$ and $2\lambda^*_i$	are the Lagrangian multipliers. Now, differentiating (\ref{Lagran}) with respect to $\boldsymbol{a}_i$ and equating to null vector $\boldsymbol{0}$ of dimension $r\times 1$, we get
\begin{equation*}
		\frac{\partial}{\partial \boldsymbol{a}_1 } \begin{vmatrix}
			W_{11} & W_{12} & \cdots & W_{1l}\\
			W_{12} & W_{22} & \cdots & W_{2l}\\
			\vdots & \vdots &  & \vdots \\
			W_{1l} & W_{2l} & \cdots & W_{ll}\\
			\end{vmatrix} -2\lambda_1\boldsymbol{1}-2\lambda^*_1\boldsymbol{\alpha} = \boldsymbol{0},
\end{equation*}
or, 
\begin{equation*}\label{G1}
 \begin{vmatrix}
			\frac{1}{2}\frac{\partial W_{11}}{\partial \boldsymbol{a}_1 } & 	\frac{\partial W_{12}}{\partial \boldsymbol{a}_1 } & \cdots & 	\frac{\partial W_{1l}}{\partial \boldsymbol{a}_1 }\\
		W_{12} & W_{22} & \cdots & W_{2l}\\
		\vdots & \vdots &  & \vdots \\
		W_{1l} & W_{2l} & \cdots & W_{ll}\\
	\end{vmatrix}+
 \begin{vmatrix}
	\frac{1}{2}\frac{\partial W_{11}}{\partial \boldsymbol{a}_1 } & W_{12} & \cdots & 	W_{1l}\\
	\frac{\partial W_{12}}{\partial \boldsymbol{a}_1 } & W_{22} & \cdots & W_{2l}\\
	\vdots & \vdots &  & \vdots \\
	\frac{\partial W_{1l}}{\partial \boldsymbol{a}_1 } & W_{2l} & \cdots & W_{ll}\\
\end{vmatrix} = 2\lambda_1\boldsymbol{1}-2\lambda^*_1\boldsymbol{\alpha},
\end{equation*}which is equivalent to
\begin{equation*}
	2\begin{vmatrix}
		\frac{1}{2} W^{(1)}_{11} & 	\frac{1}{2} W^{(2)}_{22} & \cdots & 	\frac{1}{2} W^{(l)}_{ll}\\
		W_{12} & W_{22} & \cdots & W_{2l}\\
		\vdots & \vdots &  & \vdots \\
		W_{1l} & W_{2l} & \cdots & W_{ll}\\
	\end{vmatrix}	= 2\lambda_1\boldsymbol{1}-2\lambda^*_1\boldsymbol{\alpha}.
\end{equation*}Expanding now the above determinant by its first row with corresponding co-factor terms, we obtain
\begin{equation*}
	(\boldsymbol{\Sigma}\boldsymbol{a}_1 - \boldsymbol{\omega}_{s_1})C^{11}+(\boldsymbol{\Sigma}\boldsymbol{a}_2 - \boldsymbol{\omega}_{s_2})C^{12} + \cdots+ (\boldsymbol{\Sigma}\boldsymbol{a}_l - \boldsymbol{\omega}_{s_l})C^{1l}= \lambda_1\boldsymbol{1}-\lambda^*_1\boldsymbol{\alpha}.
\end{equation*}Similarly, differentiating (\ref{Lagran}) with respect to $\boldsymbol{a}_i,~ i = 2,\cdots, l,$ and simplifying, we get
\begin{eqnarray*}
	(\boldsymbol{\Sigma}\boldsymbol{a}_1 - \boldsymbol{\omega}_{s_1})C^{12}+(\boldsymbol{\Sigma}\boldsymbol{a}_2 - \boldsymbol{\omega}_{s_2})C^{22} + \cdots+ (\boldsymbol{\Sigma}\boldsymbol{a}_l - \boldsymbol{\omega}_{s_l})C^{2l}&=& \lambda_1\boldsymbol{1}-\lambda^*_1\boldsymbol{\alpha},\\
	\vdots~~~~~~~~~~~~~~~~~~~~~~~~~~~~~&=&~~~~~~~\vdots\\
	(\boldsymbol{\Sigma}\boldsymbol{a}_1 - \boldsymbol{\omega}_{s_1})C^{1l}+(\boldsymbol{\Sigma}\boldsymbol{a}_2 - \boldsymbol{\omega}_{s_2})C^{2l} + \cdots+ (\boldsymbol{\Sigma}\boldsymbol{a}_l - \boldsymbol{\omega}_{s_l})C^{ll}&=& \lambda_1\boldsymbol{1}-\lambda^*_1\boldsymbol{\alpha}.	
\end{eqnarray*}Combining all the above $l$ equations, we can write in matrix notation 
\begin{equation}\label{Meq}
	\begin{bmatrix}
		C^{11} & C^{12} & \cdots & C^{1l}\\
		C^{12} & C^{22} & \cdots & C^{2l}\\
		\vdots & \vdots &  & \vdots \\
		C^{1l} & C^{2l} & \cdots & C^{ll}\\
	\end{bmatrix}
\begin{bmatrix}
	\boldsymbol{\Sigma}\boldsymbol{a}_1 - \boldsymbol{\omega}_{s_1}\\
	\boldsymbol{\Sigma}\boldsymbol{a}_2 - \boldsymbol{\omega}_{s_2}\\
	\vdots\\
	\boldsymbol{\Sigma}\boldsymbol{a}_l - \boldsymbol{\omega}_{s_l}
\end{bmatrix}	=
\begin{bmatrix}
	\lambda_1\boldsymbol{1}-\lambda^*_1\boldsymbol{\alpha}\\
	\lambda_2\boldsymbol{1}-\lambda^*_2\boldsymbol{\alpha}\\
	\vdots\\
	\lambda_l\boldsymbol{1}-\lambda^*_l\boldsymbol{\alpha}
\end{bmatrix}.
\end{equation}With $\boldsymbol{C} =  \big(\big(C^{ij}\big)\big)_{i, j=1}^l$ denoting the adjoint matrix of $\boldsymbol{W}$, it is known that $$\boldsymbol{C}=|\boldsymbol{W}|\boldsymbol{W}^{-1}.$$As $\boldsymbol{W}$ is positive-definite and is invertible, so is $\boldsymbol{C}$, and in fact, $$\boldsymbol{C}^{-1} = \frac{1}{|\boldsymbol{W}|}\boldsymbol{W}.$$Thence, by pre-multiplying (\ref{Meq}) by $\boldsymbol{C}^{-1}$ on both sides, we obtain simply  
\begin{equation*}
	\begin{bmatrix}
		\boldsymbol{\Sigma}\boldsymbol{a}_1 - \boldsymbol{\omega}_{s_1}\\
		\boldsymbol{\Sigma}\boldsymbol{a}_2 - \boldsymbol{\omega}_{s_2}\\
		\vdots\\
		\boldsymbol{\Sigma}\boldsymbol{a}_l - \boldsymbol{\omega}_{s_l}
	\end{bmatrix}	=  \frac{1}{|\boldsymbol{W}|} \begin{bmatrix}
	W_{11} & W_{12} & \cdots & W_{1l}\\
	W_{12} & W_{22} & \cdots & W_{2l}\\
	\vdots & \vdots &  & \vdots \\
	W_{1l} & W_{2l} & \cdots & W_{ll}\\
\end{bmatrix}
\begin{bmatrix}
	\lambda_1\boldsymbol{1}-\lambda^*_1\boldsymbol{\alpha}\\
	\lambda_2\boldsymbol{1}-\lambda^*_2\boldsymbol{\alpha}\\
	\vdots\\
	\lambda_l\boldsymbol{1}-\lambda^*_l\boldsymbol{\alpha}
\end{bmatrix}.
\end{equation*}So, the solutions for $\boldsymbol{a}_i,~ i = 1, 2,\cdots, l,$ are readily obtained from the above equation as
\begin{eqnarray}\nonumber
	\boldsymbol{a}_1 &=& \boldsymbol{\Sigma}^{-1}\boldsymbol{\omega}_{s_1} + \frac{1}{|\boldsymbol{W}|}(W_{11}\lambda_1 + W_{12}\lambda_2+\cdots+W_{1l}\lambda_l) \boldsymbol{\Sigma}^{-1} \boldsymbol{1}\\\label{a1}
	&&~~~~~~~~~~~~+\frac{1}{|\boldsymbol{W}|}(W_{11}\lambda^*_1 + W_{12}\lambda^*_2+\cdots+W_{1l}\lambda^*_l) \boldsymbol{\Sigma}^{-1} \boldsymbol{\alpha},\\\nonumber
		\boldsymbol{a}_2 &=& \boldsymbol{\Sigma}^{-1}\boldsymbol{\omega}_{s_2} + \frac{1}{|\boldsymbol{W}|}(W_{12}\lambda_1 + W_{22}\lambda_2+\cdots+W_{2l}\lambda_l) \boldsymbol{\Sigma}^{-1} \boldsymbol{1}\\\nonumber
	&&~~~~~~~~~~~~+\frac{1}{|\boldsymbol{W}|}(W_{12}\lambda^*_1 + W_{22}\lambda^*_2+\cdots+W_{2l}\lambda^*_l) \boldsymbol{\Sigma}^{-1} \boldsymbol{\alpha},\\\nonumber
\vdots~~&&	\\\nonumber
		\boldsymbol{a}_l &=& \boldsymbol{\Sigma}^{-1}\boldsymbol{\omega}_{s_l} + \frac{1}{|\boldsymbol{W}|}(W_{1l}\lambda_1 + W_{2l}\lambda_2+\cdots+W_{ll}\lambda_l) \boldsymbol{\Sigma}^{-1} \boldsymbol{1}\\\nonumber
&&~~~~~~~~~~~~+\frac{1}{|\boldsymbol{W}|}(W_{1l}\lambda^*_1 + W_{2l}\lambda^*_2+\cdots+W_{ll}\lambda^*_l) \boldsymbol{\Sigma}^{-1} \boldsymbol{\alpha}.
\end{eqnarray}As in Theorem 1, let us introduce the notation
\begin{eqnarray*}
	A_{s_i} &=& 1 - \boldsymbol{1}^{\prime}\boldsymbol{\Sigma}^{-1}\boldsymbol{\omega}_{s_i},\\
	B_{s_i} &=& \alpha_{s_i} - \boldsymbol{\alpha}^{\prime}\boldsymbol{\Sigma}^{-1}\boldsymbol{\omega}_{s_i};
	\end{eqnarray*}for $i = 1, 2,\cdots, l.$ Now, upon using the unbiasedness conditions $\boldsymbol{a}^{\prime}_i\boldsymbol{1} = 1$ and $\boldsymbol{a}^{\prime}_i\boldsymbol{\alpha} = \alpha_{s_i}$, for $i = 1, 2,\cdots, l,$ we get
\begin{eqnarray*}
	W_{11}(V_1\lambda_1 + V_3\lambda^{*}_1) + W_{12}(V_1\lambda_2 + V_3\lambda^{*}_3)+\cdots+W_{1l}(V_1\lambda_l + V_3\lambda^{*}_l) &=&  |\boldsymbol{W}|A_{s_1}, \\
	W_{11}(V_3\lambda_1 + V_2\lambda^{*}_1) + W_{12}(V_3\lambda_2 + V_2\lambda^{*}_3)+\cdots+W_{1l}(V_3\lambda_l + V_2\lambda^{*}_l) &=&  |\boldsymbol{W}|B_{s_1},\\
	\vdots~~~~~~~~~~~~~~~~~~~~~~~~~~~~~~~~~~~&=& ~~~~\vdots\\
	W_{1l}(V_1\lambda_1 + V_3\lambda^{*}_1) + W_{2l}(V_1\lambda_2 + V_3\lambda^{*}_3)+\cdots+W_{ll}(V_1\lambda_l + V_3\lambda^{*}_l) &=&  |\boldsymbol{W}|A_{s_l}, \\
	W_{1l}(V_3\lambda_1 + V_2\lambda^{*}_1) + W_{2l}(V_3\lambda_2 + V_2\lambda^{*}_3)+\cdots+W_{ll}(V_3\lambda_l + V_2\lambda^{*}_l) &=&  |\boldsymbol{W}|B_{s_l}.
\end{eqnarray*}Finally, combining all the above $2l$ equations, in matrix notation, we have
\begin{equation}\label{MatL}
	\begin{bmatrix}
	\boldsymbol{W} & \boldsymbol{0}\\
    \boldsymbol{0} & \boldsymbol{W}
	\end{bmatrix}
\begin{bmatrix}
	\boldsymbol{V}_{13}\\
	\boldsymbol{V}_{23}
\end{bmatrix}
 = |\boldsymbol{W}|\begin{bmatrix}
 	\boldsymbol{A}\\
 	\boldsymbol{B}
 \end{bmatrix},
\end{equation}where $\boldsymbol{0}$ is a null matrix of dimension $l\times l$ and $\boldsymbol{V}_{13}$, $\boldsymbol{V}_{23}$, $\boldsymbol{A}$, $\boldsymbol{B}$ are given by
\begin{equation*}
	\boldsymbol{V}_{13} = \begin{bmatrix}
		V_1\lambda_1 + V_3\lambda^{*}_1\\
		V_1\lambda_2 + V_3\lambda^{*}_2\\
		\vdots\\
		V_1\lambda_l + V_3\lambda^{*}_l\\
	\end{bmatrix},
\boldsymbol{V}_{23} = \begin{bmatrix}
	V_3\lambda_1 + V_2\lambda^{*}_1\\
	V_3\lambda_2 + V_2\lambda^{*}_2\\
	\vdots\\
	V_3\lambda_l + V_2\lambda^{*}_l
\end{bmatrix},
\boldsymbol{A} = \begin{bmatrix}
	A_{s_1}\\
	A_{s_2}\\
	\vdots\\
	A_{s_l}
\end{bmatrix},
\boldsymbol{B} = \begin{bmatrix}
	B_{s_1}\\
	B_{s_2}\\
	\vdots\\
	B_{s_l}
\end{bmatrix}.
\end{equation*}Eq. (\ref{MatL}) can be further simplified as
\begin{equation*}
	\begin{bmatrix}
		\boldsymbol{V}_{13}\\
		\boldsymbol{V}_{23}
	\end{bmatrix} = |\boldsymbol{W}| 	\begin{bmatrix}
	\boldsymbol{W}^{-1} & \boldsymbol{0}\\
	\boldsymbol{0} & \boldsymbol{W}^{-1}
\end{bmatrix}\begin{bmatrix}
\boldsymbol{A}\\
\boldsymbol{B}
\end{bmatrix},
\end{equation*}so that
\begin{equation}\label{MatL1}
	\begin{bmatrix}
		\boldsymbol{V}_{13}\\
		\boldsymbol{V}_{23}
	\end{bmatrix} = 	\begin{bmatrix}
		\boldsymbol{C} & \boldsymbol{0}\\
		\boldsymbol{0} & \boldsymbol{C}
	\end{bmatrix}\begin{bmatrix}
		\boldsymbol{A}\\
		\boldsymbol{B}
	\end{bmatrix}. 
\end{equation}Eq. (\ref{MatL1}) yields the solutions for $\lambda_i$ and $\lambda^*_i$, ~$i = 1, 2,\cdots, l,$ as
\begin{eqnarray}\label{lambda1}
	\begin{bmatrix}
		\lambda_1\\
		\lambda_2\\
		\vdots\\
		\lambda_l
	\end{bmatrix} &=& \frac{1}{\Delta} \begin{bmatrix}
	C^{11} & C^{12} & \cdots & C^{1l}\\
	C^{12} & C^{22} & \cdots & C^{2l}\\
	\vdots & \vdots &  & \vdots \\
	C^{1l} & C^{2l} & \cdots & C^{ll}\\
\end{bmatrix} 
\begin{bmatrix}
	A_{s_1}V_2 - B_{s_1}V_3\\
	A_{s_2}V_2 - B_{s_2}V_3\\
	\vdots\\
	A_{s_l}V_2 - B_{s_l}V_3\\
\end{bmatrix},\\\label{lambda2}
\begin{bmatrix}
	\lambda^*_1\\
	\lambda^*_2\\
	\vdots\\
	\lambda^*_l
\end{bmatrix} &=& \frac{1}{\Delta} \begin{bmatrix}
	C^{11} & C^{12} & \cdots & C^{1l}\\
	C^{12} & C^{22} & \cdots & C^{2l}\\
	\vdots & \vdots &  & \vdots \\
	C^{1l} & C^{2l} & \cdots & C^{ll}\\
\end{bmatrix} 
\begin{bmatrix}
	B_{s_1}V_1 - A_{s_1}V_3\\
	B_{s_2}V_1 - A_{s_2}V_3\\
	\vdots\\
	B_{s_l}V_1 - A_{s_l}V_3\\
\end{bmatrix}.
\end{eqnarray}
Now, the coefficients of $\boldsymbol{\Sigma}^{-1} \boldsymbol{1}$ and $\boldsymbol{\Sigma}^{-1} \boldsymbol{\alpha}$ in (\ref{a1}), which can be simplified using (\ref{lambda1}) and (\ref{lambda2}), are given by
\begin{eqnarray}\label{k1}
	W_{11}\lambda_1 + W_{12}\lambda_2+\cdots+W_{1l}\lambda_l&=& \frac{|\boldsymbol{W}| }{\Delta}(A_{s_1}V_2 - B_{s_1}V_3),\\\label{k2}
	W_{11}\lambda^*_1 + W_{12}\lambda^*_2+\cdots+W_{1l}\lambda^*_l&=& \frac{|\boldsymbol{W}| }{\Delta}(B_{s_1}V_1 - A_{s_1}V_3).
\end{eqnarray}By using (\ref{k1}) and (\ref{k2}) in (\ref{a1}), the expression for  $\boldsymbol{a}_1$ becomes
\begin{eqnarray*}
	\boldsymbol{a}_1 &=& \boldsymbol{\Sigma}^{-1}\boldsymbol{\omega}_{s_1} + \frac{1}{\Delta}(A_{s_1}V_2 - B_{s_1}V_3)\boldsymbol{\Sigma}^{-1} \boldsymbol{1} + \frac{1}{\Delta}(B_{s_1}V_1 - A_{s_1}V_3)\boldsymbol{\Sigma}^{-1} \boldsymbol{\alpha}.
\end{eqnarray*} 
 In a similar manner, it can be shown that
 \begin{eqnarray*}
 	\boldsymbol{a}_2 &=& \boldsymbol{\Sigma}^{-1}\boldsymbol{\omega}_{s_2} + \frac{1}{\Delta}(A_{s_2}V_2 - B_{s_2}V_3)\boldsymbol{\Sigma}^{-1} \boldsymbol{1} + \frac{1}{\Delta}(B_{s_2}V_1 - A_{s_2}V_3)\boldsymbol{\Sigma}^{-1} \boldsymbol{\alpha},\\
 	\vdots~~&&\\
 		\boldsymbol{a}_l &=& \boldsymbol{\Sigma}^{-1}\boldsymbol{\omega}_{s_l} + \frac{1}{\Delta}(A_{s_l}V_2 - B_{s_l}V_3)\boldsymbol{\Sigma}^{-1} \boldsymbol{1} + \frac{1}{\Delta}(B_{s_l}V_1 - A_{s_l}V_3)\boldsymbol{\Sigma}^{-1} \boldsymbol{\alpha}.
 \end{eqnarray*}
These expressions of $\boldsymbol{a}^{,}_i$s correspond to the marginal BLUPs, thus completing the proof of the theorem.

\begin{cor}
	As done in Section 3, we can easily establish that the simultaneous BLUPs derived in Theorem 3 also possess the complete MSPE matrix dominance property. 
\end{cor}

\subsection{Numerical illustration}
\paragraph{}
This example uses a normally distributed Type-II right censored data, considered earlier by \cite{Doganaksoy_1997}. A life-test with $n=10$ resulted in a Type-II censored sample of $r=5$ failure times as 87.0, 92.8, 117.1, 133.6 and 138.6.  The simultaneous BLUPs  of $(X_{6:10}, X_{7:10}, X_{8:10}, X_{9:10},$ $ X_{10:10})^{\prime}$ in this case are
\begin{equation*}
	\begin{pmatrix}
		\tilde{X}_{6:10} \\
		\tilde{X}_{7:10} \\
		\tilde{X}_{8:10}\\
		\tilde{X}_{9:10}\\
		\tilde{X}_{10:10}
	\end{pmatrix}=
\begin{pmatrix}
148.73\\
158.99\\
170.36\\
184.37\\
206.19
\end{pmatrix},
\end{equation*}which are exactly the same as their marginal BLUPs. In addition, the associated MSPE matrix is 
\begin{equation*}
\sigma^2	\begin{pmatrix}
		0.0655 & 0.0762 & 0.0885 & 0.1042 & 0.1292\\
		0.0762  & 0.1552 &  0.1785  & 0.2084 & 0.2568\\
		0.0885  &  0.1785& 0.2835  &  0.3268  &  0.3979\\
		0.1042   &  0.2083  &  0.3268  & 0.4897 &  0.5835\\
		0.1292   & 0.2568 & 0.3979 & 0.5835 & 0.9471
		
\end{pmatrix}.	
\end{equation*}

\section{Concluding remarks}
\paragraph{}
In this article, we have derived explicit expressions for the simultaneous BLUPs of any $l$ future order statistics. We have specifically shown that the simultaneous BLUPs are the same as the marginal BLUPs. It is important to keep in mind that the BLUPs based on the determinant of the variance–covariance matrix of the predictors do not exist for $l$ larger than 2 \citep[see][]{Bala_2021}. Thus, as an alternative, the simultaneous BLUPs based on the minimization of MSPE matrix has an advantage. We are currently looking into the simultaneous best linear invariant prediction issue and hope to report the findings in a future paper.

\bibliographystyle{apalike}
\bibliography{ritwik_ref}

\end{document}